\documentclass[12pt,tightenlines,eqsecnum,floats,showpacs,nofootinbib,amsmath,amssymb,aps,prd]{revtex4}

\usepackage{setspace}
\usepackage{amsmath,amssymb,amsfonts,amsthm,latexsym}
\usepackage{graphicx}
\usepackage{enumerate} 
\usepackage{colordvi} 

\def\be{\begin{equation}}
\def\ee{\end{equation}}
\def\ba{\begin{eqnarray}}
\def\ea{\end{eqnarray}}
\def\f{\frac}

\def\h{\hat}

\def\v{\vec}

\def\I{\mathbb{I}}
\def\L{\Lambda}
\def\H{\mathcal{H}}

\def\H{{\mathcal{H}}}

\def\dd{{\rm d}}
\def\WDW{\rm WDW\,\,}

\def\a{\mathfrak{a}}
\def\W{\mathfrak{W}}

\def\SU(2){\rm SU(2)}

\newcommand{\bra}{\langle}
\newcommand{\ket}{\rangle}

\newcommand{\R}{{\mathbb{R}}}
\newcommand{\C}{{\mathbb{C}}}
\newcommand{\E}{E}

\begin{document}


\title{Some surprising implications of background independence\\ in
canonical quantum gravity}
\author{Abhay Ashtekar}
\affiliation{Institute for Gravitation and the Cosmos, \& Physics
Department, Penn State, University Park, PA 16802, U.S.A.}

\begin{abstract}

There is a precise sense in which the requirement of background
independence suffices to uniquely select the kinematics of loop
quantum gravity (LQG). Specifically, the fundamental kinematic
algebra of LQG admits a \emph{unique} diffeomorphism invariant
state. Although this result has been established rigorously, it
comes as a surprise to researchers working with other approaches to
quantum gravity. The goal of this article is to explain the
underlying reasons in a pedagogical fashion using geometrodynamics,
keeping the technicalities at their minimum. This discussion will
bring out the surprisingly powerful role played by diffeomorphism
invariance (and covariance) in non-perturbative, canonical quantum
gravity.

\end{abstract}

\maketitle
\section{Introduction}
\label{s1}

It is a pleasure and an honor to dedicate this article to the memory
of Professor J\"urgen Ehlers, a model teacher, mentor and friend. I
was fortunate to have had him as a teacher already for my first
graduate course in cosmology and he continued to educate me on
almost all aspects of theoretical physics, including mathematical
physics, quantum field theory, foundations of quantum mechanics,
statistical physics and of course innumerable topics in general
relativity. I deeply admired many traits that were unique to him.
The one that I miss most is his unmatched conceptual clarity that
led him to offer clear explanations that went straight to the heart
of the matter. In this homage therefore I will attempt to clarify a
rather fundamental issue that he would have appreciated: the
surprisingly powerful role played by diffeomorphism invariance in
non-perturbative quantum gravity.

Let me begin by explaining a key mathematical result that appears
puzzling to anyone who hears it for the first time. For brevity, I
will just summarize the basic of the picture here; details will
follow in subsequent sections.

In loop quantum gravity (LQG), the basic canonical fields are an
$\SU(2)$ connection $A_a^i$ and the conjugate `electric 2-forms'
$\E_{ab}^i$ on a 3-manifold $M$.%
\footnote{Although I will restrict myself to 3 dimensions for
definiteness, all considerations in this paper will go through in
higher dimensions as well. Indeed the fundamental uniqueness results
\cite{lost,cf} I am about to explain were established for all
dimensions greater than 1. For simplicity of discussion, I will
assume that $M$ is compact but this restriction can be removed by
paying due attention to boundary conditions.}
Because we are dealing with gravity, the electric 2-forms have a
dual interpretation: they determine a Riemannian geometry on $M$,
the details of which are not important for our purpose. The Pairs
$(A_a^i, \E_{ab}^i)$ coordinatize the gravitational phase space. The
`elementary' variables are taken to be the holonomies $h_e(A)$ of
connections $A_a^i$ along suitably chosen edges $e$ and fluxes of
electric fields $\E_{S,f}:= \int_S f_i \E^i$, across suitably chosen
2-surfaces $S$ in $M$ (where $f_i$ is a test field on $S$ which
takes values in the dual of the Lie algebra ${\rm su(2)}$). These
holonomies $h_e$ and electric fluxes $\E_{S,f}$ generate the
kinematical $\star$-algebra $\a$ of quantum operators.

Note that the algebra is constructed abstractly; a priori there is
no Hilbert space for it to act upon. In non-relativistic quantum
mechanics, one generally introduces the Hilbert space of states $\H$
first and then 
defines operators by specifying their action on elements of $\H$.
This is because, subject to certain well-motivated technical
conditions, the (abstractly defined) Heisenberg algebra generated by
the position and momentum operators ${q}$ and ${p}$
---or the Weyl algebra generated by their exponentials
$W(\lambda,\,\mu) = e^{i\lambda q + \mu p}$
where $\lambda,\mu$ are real parameters with appropriate physical
dimensions--- admits a unique irreducible representation (up to
unitary equivalence). This celebrated von-Neumann uniqueness theorem
enables us to restrict ourselves 
just to the Schr\"odinger representation. For systems with an
infinite number of degrees of freedom, on the other hand, the
uniqueness fails and there are infinitely many representations of
the fundamental operator algebras \cite{gw,rh}. The prevalent
philosophy in the mathematical literature is that it is dynamics
that selects which representation one must use. Therefore one first
constructs the algebra abstractly, without reference to a Hilbert
space (see, e.g., \cite{ge}).

A \emph{state} on a $\star$-algebra $\a$ is a positive linear
functional (PLF), i.e., a linear map $F: \, \a \rightarrow \C$
\,\,such that $F[a^\star a] \ge 0$. If $\a$ is represented as an
algebra of operators on a Hilbert space $\H$, every element $\Psi
\in \H$ defines a PLF $F_\Psi$ on $\a$ through expectation values:
$F_\Psi[A] = \bra\Psi,\, A \Psi\ket$.  As we will see in section
\ref{s2}, every state on an abstract $\star$ algebra naturally leads
to a Hilbert space and a representation of elements of $\a$ by
operators on that Hilbert space. This is the celebrated Gel'fand,
Naimark, Segal (GNS) construction. In quantum theory of free fields,
for example, one can uniquely select a PLF $F_o$ by making appeal to
general physical principles ---in particular the Poincar\'e
invariance--- and the GNS construction then yields the Fock
representation in which $F_o$ is given by the vacuum expectation
values.

A deep mathematical result due to Lewandowski, Okolow, Sahlmann and
Thiemann \cite{lost} is that, subject to certain well-motivated
technical assumptions, the $\star$-algebra $\a$ of LQG \emph{admits
a unique diffeomorphism invariant state.} Fleischhack \cite{cf} has
proved another uniqueness theorem. He works with the Weyl algebra
$\W$ generated by the holonomies and the exponentials of the
electric flux operators and his technical conditions are somewhat
different from those used in \cite{lost}. The two results are
complementary, each with its relative technical advantages. Although
as rigorous theorems they are distinct, at my rather general level
of discussion the two have the same physical content. The GNS
construction then provides a representation of $\a$ which
constitutes the basic kinematical framework of LQG. This is the
technical arena for first formulating and then solving the quantum
constraints (which encode quantum dynamics). This arena was
introduced and developed already in the 90's; it leads to a specific
quantum theory of Riemannian geometry
\cite{alrev,crbook,ttbook,almmt,al5,rs,alvol,acz,tt}. Although the
uniqueness results of Lewandowski, Okolow, Sahlmann, Thiemann and
Fleischhack came several years after the main kinematical results of
LQG, they provide the logical foundation for quantum geometry.

Yet, when one takes a moment to think about it, this central result
seems very puzzling: How can there be an \emph{unique}
diffeomorphism state? Surely, quantum gravity admits an infinite
number of diffeomorphism invariant states! Consider either the older
Wheeler-DeWitt (WDW) theory \cite{jw,kk} based on 3-metrics $q_{ab}$
and the conjugate momenta $p^{ab}$, or, Klauder's more recent affine
quantum gravity \cite{jk}. In these approaches, one can represent
states as functionals $\Psi(q)$ of 3-metrics. Surely there is an
infinite number of diffeomorphism invariant wave functions
$\Psi(q)$: one can, for example, just integrate any curvature
invariant constructed from $q_{ab}$ over $M$. How does one reconcile
this `obvious fact' with the uniqueness result of \cite{lost,cf}?

The first goal of this article is to show, by a careful analysis of
the \WDW theory, that there is in fact no tension. We will see that
there are apparently surprising results also in the \WDW theory;
diffeomorphism invariance is a much stronger requirement that one
might have first thought. Results can seem counter intuitive if one
does not carefully distinguish between the kinematical algebra and
the algebra of diffeomorphism invariant variables. The second goal
is to provide intuition for the origin of the two features of the
kinematics of LQG which are not shared by 
familiar Minkowskian quantum field theories ---the non-separability
of the kinematical Hilbert space and the non-existence of a local
connection operator $A_a^i(x)$. We will see that they can be traced
back to the diffeomorphism covariance of LQG. Finally, it will be
instructive to compare the role of gauge invariance in the Maxwell
theory with that of diffeomorphism invariance. We will find that,
because of its inherent non-locality, diffeomorphism invariance is
much more powerful than gauge invariance.

The article is organized as follows. In section \ref{s2}, I recall
the GNS construction and show how a diffeomorphism invariant state
leads to a representation of the algebra on which the diffeomorphism
group is unitarily represented. This is the ideal setting for
solving the diffeomorphism constraint via group averaging
\cite{dm,almmt}. In section \ref{s3}, I consider geometrodynamics a
la \WDW and show that, contrary to one's first expectation, the
Heisenberg algebra \emph{admits no diffeomorphism invariant state}.%
\footnote{We will also show that the affine algebra of Klauder's
\cite{jk} admits such a state but the resulting GNS representation
turns out to be trivial.}
The Weyl algebra does and the most natural such state is completely
analogous to the one used in LQG. The GNS construction again leads
to a non-separable Hilbert space and, furthermore, does not admit a
local momentum operator ${p}_{ab}$. So, the situation is completely
analogous to that in LQG. Section \ref{s4} is devoted to a summary
and a discussion of the difference between the gauge group of the
Maxwell theory and the diffeomorphism group of gravity.

Since I hope to reach a diverse audience, I will keep the
technicalities at their minimum. Inevitably, then, I will have to
gloss over some of the finer points and ignore some mathematical
subtleties. But the essence of the argument has not been simplified.

\section{The GNS construction} \label{s2}

Let us begin with some preliminaries. A \emph{$\star$-algebra} $\a$
is an associative algebra (over the field of complex numbers)
equipped with an involution operation, denoted $\star$. By
definition, this operation satisfies the following three relations:
i) $(A+ \lambda B)^\star = A^\star + \bar\lambda B^\star$; ii)
$(AB)^\star = B^\star A^\star$; and, iii) $(A^\star)^\star = A$ for
all $A,B \in \a$ and complex numbers $\lambda$, where $\bar\lambda$
is the complex conjugate of $\lambda$. \emph{Throughout the rest of
the paper we assume that the $\star$ algebra has an identity
element.} A familiar example is the $\star$-algebra $\a$ generated
by the position and momentum operators on the Hilbert space $\H =
L^2(\R, dx)$ of quantum mechanics, where the $\star$-operation is
given by the Hermitian adjoint
on $\H$.%
\footnote{This $\a$ will consist of finite linear combinations of
finite products of ${q}$ and ${p}$. Technically it is easiest to
define all these operators on the common dense domain $C^\infty_0$
of smooth functions $\Psi(x)$ of compact support.}
In the algebraic approach, one first defines the $\star$-algebra
abstractly, without reference to a Hilbert space and then looks of
its representations. The GNS construction provides a natural tool
for this task.

An \emph{automorphism} $\theta$ on a $\star$ algebra $\a$ is a
linear map from $\a$ onto itself which preserves the product and the
$\star$-operation, i.e., satisfies $\theta(AB) = \theta(A)
\,\theta(B)$ and $\theta(A^\star) = (\theta(A))^\star$, for all $A,B
\in \a$. Thus, $\theta$ preserves the structure of $\a$ and
therefore represents a symmetry. If $\a$ is the $\star$-algebra of
all bounded operators on a Hilbert space $\H$, every unitary map $U$
on $\H$ defines an automorphism $\theta$ on $\a$ via: $\theta(A) :=
U^{-1} A U$ for all $A \in \a$. Such `unitarily implementable'
automorphisms $\theta$ are called \emph{inner automorphisms}.

We can now summarize the GNS construction which is a powerful
technique to find representations of $\star$-algebras $\a$ which is
furthermore tailored to realize symmetries $\theta$ of $\a$ by
unitary operators on the Hilbert space defined in the
representation.

Fix a $\star$-algebra $\a$ with an identity element and a PLF $F$
thereon. We will collect a few useful properties of $F$. First, note
that since $F[A^\star A] \ge 0$ for all $A \in \a$, substituting
$A^\star$ for $A$, we immediately conclude $F[A A^\star] \ge 0$ for
all $A \in \a$. Second, using the fact that $F[(A+\lambda B)^\star
(A+\lambda B)] \ge 0$ for all $A, B \in \a$ and $\lambda \in \C$, it
immediately follows that:
\be \label{1}F[AB^\star] = \bar{F}[BA^\star]\quad \hbox{\rm and
hence in particular}\quad F[A^\star] = \bar{F}[A]  \ee
and
\be \label{2} |F[A^\star B]|  \le 2 \sqrt{F[A^\star A] F[B^\star
B]}\,  \ee
for all $A, B$ in $\a$, where the over-bar denotes complex
conjugation.

We are now ready to construct the representation defined by $F$. To
explain the central idea, let me for a moment assume that $F$ is
strictly positive, i.e., that $F[A^\star A] \ge 0$ where the
equality holds if and only of $A=0$ and remove this assumption in a
second step. Then, we can easily construct the required Hilbert
space as follows. Recall that $\a$ is in particular a complex vector
space and for clarity of notation denote this vector space by $V$
and its elements by $\vec{A}$ with an over-arrow. Define an inner
product on $V$ using $F$: $\bra \v{A},\,\, \v{B}\ket = F[A^\star
B]$. It is straightforward to check that this is an Hermitian inner
product. The required Hilbert space $\H$ is obtained by Cauchy
completion of this inner product space. Finally, we need to
represent elements $A$ of $\a$ as concrete operators $\L(A)$ on $V$,
where $\L$ denotes the representation map. Again there is natural
map: \,\, $\L(A)|\v{B}\ket = |\overrightarrow{AB}\ket$. (These
concrete operators $\L(A)$ have a common dense domain, namely $V$.)
It is straightforward to check that this is a representation, i.e.,
that $\L(A+\lambda B) = \L(A) + \lambda\, \L(B),\, \L(A^\star)=
[\L(A)]^\dag, \,$ and $\L(AB) = \L(A)\L(B)$, for all $A,B$ in $\a$
and complex numbers $\lambda$ where $\dag$ denotes the concrete
Hermitian-adjoint operation on $\H$.

Thus, the underlying idea of obtaining a $\star$-representation of
$\a$ using the state $F$ is simple and natural. However, in
practice, states are not strictly positive definite and therefore
one has to tweak this construction.  Let us now suppose that $F$ is
a general PLF. Then consider the space $K$ of operators $A \in A$
such that $F[A^\star A] =0$. Using (\ref{1}) and (\ref{2}), it is
straightforward to show that $K$ is a vector sub-space of $\a$:
\be \label{3} A, B \in K \quad \Rightarrow \quad A+\lambda B \in K\,
, \ee
and, in fact, a left ideal of $\a$:
\be \label{4} A \in K,\,\, B\in \a \quad \Rightarrow \quad BA \in
K\, . \ee
Let us now consider the complex vector space $V := \a/K$. An element
$\v{A}$ of $V$ is now \emph{an equivalence class} of operators in
$\a$, where  $\v{A}_1 = \v{A}_2$ if and only if
$F[(A_1-A_2)^\star\,(A_1-A_2)] =0$. We can now define an Hermitian
inner product on $V$, construct the Hilbert space $\H$ by Cauchy
completion of $V$ and define the representation map $\L$ exactly as
before. The only additional step required is to check that the inner
product $\bra \v{A},\, \v{B}\ket$ and the action $\L(C)|\v{A}\ket$
of concrete operators $\L(C)$ on $|\v{A}\ket$ remain unchanged if we
replace $A$ or $B$ by another operator in the equivalence classes
$\v{A}$ and $\v{B}$ respectively. However, this is straightforward
using (\ref{1})-(\ref{4}). Again, all operators have a common dense
domain, $V$. Finally, note that $K$ is \emph{not} a right ideal.
This asymmetry comes  because by definition $A\in \a$ if and only if
$F[A^\star A] = 0$, and not $F[A A^\star] =0$, and the first
condition does not imply the second for individual operators $A$.
This is just as one would expect; had $K$ been a 2-sided ideal, $V$
would have inherited a product structure from $\a$ and we know that
elements of $\H$ cannot be multiplied.

Let us denote the ket $|\v{\mathbb{I}}\ket$ in $\H$ by $|\Psi_F\ket$
where $\mathbb{I}$ is the identity operator in $\a$. It is immediate
that $|\Psi_F\ket$ is a \emph{cyclic vector}, i.e., repeated actions
of the concrete operators $\Lambda(A)$ on $|\Psi_F\ket$ generates a
dense sub-space of $\H$ (namely, $V$). Furthermore, the given PLF
$F$ is now realized as the expectation value functional in this ket:
\be \label{vev} F[A]\, =\, \bra \Psi_F,\, \L(A)\Psi_F\ket\, . \ee
In the mathematical literature $|\Psi_F\ket$ is often referred to as
the ``vacuum'' and $F[A]$ as the ``vacuum expectation value
functional,'' although from a physics perspective this is somewhat
of a misnomer since the construction is purely kinematical and does
not refer to a Hamiltonian (or a number) operator. (This is rather
analogous to the fact that in symplectic geometry the flow generated
by any function on the phase space is referred to as a Hamiltonian
flow; the function does not have to be the physical Hamiltonian of
the system.)

Next, this representation $(\H, \L)$ of $\a$ is unique in the
following sense. Let $(\H',\L')$ be another cyclic representation
with a cyclic vector $\Psi'$ such that $F(A) = \bra \Psi',\, \L'(A)
\Psi' \ket$ for all $A \in \a$. Then, there is a unitary map
$\mathbb{U}$ from $\H'$ to $\H$ such that $\L'(A)= \mathbb{U}^{-1}
\L(A) \mathbb{U}$ (and \,$\mathbb{U}\, |\Psi'\ket = |\Psi_F\ket$
and\,). In this sense the GNS representation is completely
characterized by the given PLF $F$.

Finally, suppose that we are given a 1-1 automorphism $\theta$ on
$\a$ and the given state $F$ is invariant under its action:
\be F[\theta(A)] = F[A], \quad \forall A \in \a \, .\ee
Then, we can define an unitary operator $U$ on $\H$ via $U|\v{A}\ket
= |\overrightarrow{\theta(A)}\ket$. It is easy to verify that
\be  \L\big(\theta(B)\big) = U\, \L(B)\, U^{-1}\quad \forall B\in
\a\, ;\ee
$\theta$ is thus unitarily implemented in the GNS representation of
$\a$ generated by $F$. In the quantum theory of fields in Minkowski
space, for example, the Poincar\'e invariance of the vacuum
expectation values implies that the Poincar\'e group is represented
by unitary operators on the Fock space.

To summarize, it is natural to regard the operator algebra $\a$ as
more fundamental, especially while dealing with systems with an
infinite number of degrees of freedom \cite{ge}. A Hilbert space
$\H$ can then be recovered by specifying a quantum state ---i.e., a
PLF---on $\a$. The PLF itself is realized as the expectation value
functional in the cyclic state. If the PLF is invariant under a 1-1
automorphism $\theta$, then this symmetry is unitarily represented
on the \emph{full} Hilbert space. The power of the GNS construction
lies in its generality. It encompasses all physical systems:
non-relativistic quantum mechanics, quantum fields in flat and
curved space-times, statistical systems and, as we will see in the
next two sections, diffeomorphism invariant systems such as general
relativity. Technically it is very simple because it is purely
algebraic. One does not need to use topology (and, in particular,
the $C^\star$ algebra machinery) anywhere.

In the next section we will apply these ideas to canonical quantum
gravity.

\section{Kinematics of Geometrodynamics and diffeomorphisms}
\label{s3}

We are now ready discuss the kinematics of the \WDW theory, in
particular the issue of diffeomorphism invariant states on the
kinematic algebra. To distinguish operators from test fields used to
smear them, I will generally put `hats' on operators. This section
is divided into three parts. In the first two parts I will consider
the Heisenberg algebra and in the third, the Weyl algebra.

\subsection{Heisenberg algebra of the WDW theory}
\label{s3.1}

The canonical algebra $\a$ is generally constructed as follows. The
basic objects are (abstractly defined) tensor-valued distributions
$\h{q}_{ab}, p^{ab}$ on the 3-manifold $M$ where $\h{q}_{ab}$
represents the metric and $\h{p}^{ab}$ the conjugate momentum (so
$\h{p}_{ab}$ carries a density weight 1). Thus, associated with each
real, symmetric, smooth tensor density $f^{ab}$ of weight 1 on $M$,
there is a configuration operator $\hat{q}(f) := \int_M
\h{q}_{ab}f^{ab} \dd^3x$ and with each real, symmetric, smooth
tensor density $h_{ab}$ there is a momentum operator $\hat{p}(h) :=
\int_M \h{p}^{ab}h_{ab} \dd^3x$. The fundamental $\star$-algebra
$\a$ is generated by finite linear combinations of finite products
of $\h{q}(f)$ and $\h{p}(h)$ (together with the identity operator),
subject to the $\star$-relations:
\be \label{star1}[\h{q}(f)]^\star\, =\, \h{q}(f), \quad {\rm and}
\quad [\h{p}(h)]^\star\, =\, \h{p}(h)\, ,  \ee
and the canonical commutation relations
\be \label{ccr} [\h{q}(f_1),\, \h{q}(f_2)] =0, \quad [\h{q}(f),\,\,
\h{p}(h)]\, = i\hbar\, \textstyle{\int}_M\, f^{ab}h_{ab}\,
{\dd}^3x\, ,\quad [\h{p}(h_1),\, \h{p}(h_2)] =0\, . \ee
Note that all the integrals in this discussion are well defined
without the need of any background metric or volume element because
the integrands are scalar densities of weight 1. This canonical
$\star$ algebra $\a$ constitutes the point of departure for the \WDW
theory. Following the standard literature on the \WDW theory
\cite{jw,kk}, we have not required $\h{q}_{ab}$ to be a posi\-tive
definite here. This additional condition is imposed in Klauder's
approach \cite{jk} where the canonical commutation relations have
then to be modified to affine commutation relations. We will return
to is point in section \ref{s3.2}.

Now, each smooth diffeomorphisms $\alpha$ on the underlying
3-manifold $M$ defines a natural automorphism $\theta_\alpha$ on
$\a$ via:
\be \theta_\alpha (\h{q}(f)) = \h{q}(\v\alpha (f)), \quad {\rm and}
\quad \theta_\alpha (\h{p}(h)) = \h{p}(\underleftarrow\alpha (h))\ee
where $\vec{\alpha}(f)$ is the push-forward of $f^{ab}$ under the
diffeomorphism $\alpha$ and $\underleftarrow\alpha(h)$, the
pull-back of $h_{ab}$ under $\alpha$.

We can now investigate diffeomorphism invariant states on $\a$.
Since $\a$ is generated by operator-valued distributions
$\h{q}_{ab}$ and $\h{p}^{ab}$, following the standard procedure in
Minkowskian quantum field theories, it is natural to represent
states by PLFs $F$ for which $F[\h{q}_{ab}]$ is a tensor valued
distribution and $F[\h{p}^{ab}]$ is a tensor-valued distribution
with density weight 1. More precisely, one requires that
$F[\h{q}(f)], F[\h{q}(f_1)\h{q}(f_2)$, \ldots  $F[\h{q}(f_1) \ldots
\h{q}(f_n)]$ and $F[\h{p}(h)], F[\h{p}(h_1)\h{p}(h_2)$, \ldots
$F[\h{p}(h_1) \ldots \h{p}(h_n)]$ are continuous maps from the space
of test fields to the reals (in the standard topology on the space
of test fields). A state $F$ on $\a$ is diffeomorphism invariant if
and only if
\be\label{diffinv} F[\theta_\alpha(A)]  = F[A] \ee
for all smooth diffeomorphisms $\alpha$ on $M$ and all $A$ in $\a$.
The question for us is: How many diffeomorphism invariant states
does $\a$ admit?

Let $F$ be such a state. Then, there exists a distribution
$Q_{ab}(x)$ on $M$ such that $F[\h{q}(f)] = \int_M Q_{ab}(x)
f^{ab}(x)\, \dd^3x$. Similarly, $F[\h{q}(f_1) \h{q}(f_2)]$ defines a
bi-distribution $Q_{ab\,a'b'}(x,x')$. The requirement
(\ref{diffinv}) of diffeomorphism invariance implies that these
distributions are all diffeomorphism invariant. But the only
tensor-valued (multi-)distributions $G_{ab}(x),\,
G_{ab\,a'b'}(x,x'),\, \ldots$ which are diffeomorphism invariant are
the zero distributions.%
\footnote{Here and in the rest of this sub-section, the fact that
all indices on these distributions are covariant (or contravariant)
and they have certain density weights is important. There do exist
non-trivial diffeomorphism invariant tensor distributions with
certain index structures and density weights. An example is
$T_{ab}{}^{a'b'}(x,x') = \delta_a^{a'}\, \delta_b{}^{b'}\,
\delta(x,x')$. But the ones we consider are not of this type.}
Let us now consider the GNS representation determined by $F$. Since
$F[\h{q}(f)]=0$,\, $F[\h{q}(f_1) \h{q}(f_2)]=0,\, \ldots$ it follows
that
\be \bra \Psi_F,\,\L(\h{q}(f))\,\Psi_F\ket = 0, \quad \bra \Psi_F,\,
\L(\h{q}(f_1)\h{q}(f_2))\, \Psi_F\ket =0, \quad \ldots \, .\ee
for all $f, f_1, f_2, \ldots$. This implies in particular that
$\L(\h{q}(f))|\Psi_F\ket =0$ for all $f$. Identical argument implies
$\L(\h{p}(h))|\Psi_F\ket =0$ for all $h$. But this implies
\be \label{contra}  \big( \L(\h{q}(f))\,
\L(\h{p}(h))\big)|\Psi_F\ket = 0, \quad {\rm and} \quad  \big(
\L(\h{p}(h))\, \L(\h{q}(f))\big)\,|\Psi_F\ket = 0\, . \ee
Now, if a non-zero, diffeomorphism invariant PLF $F$ had existed, as
we saw in section \ref{s2}, the GNS construction would have
guaranteed that $\L$ is a representation of $\a$ and, in particular,
$\big(\L(\h{q}(f))\, \L(\h{p}(h))\,-\, \L(\h{p}(h))\,
\L(\h{q}(f))\big) \Psi_F = i\hbar (\int_M f^{ab}(x) t_{ab}(x)
\dd^3x) |\Psi_F\ket$. Thus, (\ref{contra}) contradicts the canonical
commutation relations (\ref{ccr}) whence \emph{our assumption that
there is a diffeomorphism invariant state on $\a$ cannot hold.} This
is in striking contrast to the initial expectation that there should
be an infinite number of such states.

I want to emphasize that the entire argument is kinematical and
independent of the details of dynamics. Thus, it is applicable well
beyond general relativity (and also in higher dimensions). However,
the non-existence of diffeomorphism invariant states on $\a$ does
\emph{not} mean that $\a$ cannot admit interesting representations.
But the construction will involve some structure that will break
diffeomorphism invariance. In particular, such representations
cannot arise from diffeomorphism invariant measures on the space of
metrics or a suitable completion thereof. As a consequence the
analysis of this sub-section does contradict the heuristics and
folklore which have motivated many ideas in the \WDW theory over the
years.

\subsection{Heisenberg Algebra of the affine theory} \label{s3.2}

One might first think that the problem arose because we dropped the
positivity requirement on the 3-metric $q_{ab}$. Indeed the standard
\WDW algebra allows not just degenerate metrics ---which could be
considered as limiting cases of the positive definite ones--- but
metrics of \emph{any signature}. Furthermore, even degeneracy is an
obstacle if one wishes to construct (unambiguous) curvature
invariants, which, when integrated over $M$ would provide the
intuitively obvious diffeomorphism invariant wave functions
$\Psi(q)$ mentioned in section \ref{s1}. At first, incorporation of
positivity seems difficult. But already in the 70's, Klauder
\cite{jk} showed that this could be accomplished by replacing the
canonical commutation relations (CCR) by affine commutation
relations (ACR). In this sub-section, I will show that the new
algebra also does not admit non-trivial diffeomorphism invariant
states.

The necessity of ACR can be motivated by a simple example in
non-relativistic quantum mechanics: particle confined to the
positive half of the real line. Consider \emph{any} representation
of the CCR $[q,\, p] = i\hbar$ on which $q,p$ are self-adjoint
operators. Then, $U(\mu) := \exp i\mu p$ is a 1-parameter family of
unitary operators. Since $U^{-1} q U = q - \mu\hbar$, it follows
that the spectrum of $q$ cannot be positive. Thus, there is a
tension between the CCR and the positivity requirement on $q$.
Geometrically, this can be understood as follows. On the classical
phase space the Hamiltonian vector field of the momentum function
$p$ is simply $\partial/\partial q$ and the integral curves of this
vector field do not leave the positive half of the real line
invariant. However, the Hamiltonian vector field of the function
$qp$ does map the positive half line to itself. Therefore, in the
quantum theory, we are led to consider the operators $q, \pi:=
(qp+pq)/2$ as fundamental and use their commutation relation $[q,
\pi] = i\hbar q$. These are the affine commutation relations. Now,
we can indeed represent the $\star$-algebra generated by $q$ and
$\pi$ on the Hilbert space $\H^+ := L^2(\R^+, \dd q)$ based on just
the positive half of the q-line.

In the gravitational case, then, the algebra $\a$ is now generated
by operator valued distributions $\h{q}_{ab}$ and $\h{\pi}_a{}^b$,
where we can intuitively think of $\h\pi_a{}^b$ as the quantum
analog of the function $\pi_a{}^b= q_{ac}p^{cb}$ on the classical
phase space. Thus, the $\star$-algebra $\a$ is now obtained by
considering finite sums of finite products of smeared operators
$\h{q}(f)$ and $\h\pi(g)$ (and the identity operator $\I$), where
$f^{ab}$ is as before a real, symmetric, smooth  tensor density of
weight 1 and $g^a{}_b$ is now a real, smooth tensor field on $M$.
The algebra $\a$ is subject to the $\star$-relations:
\be \label{star2}[\h{q}(f)]^\star\, =\, \h{q}(f), \quad {\rm and}
\quad [\h{\pi}(g)]^\star\, =\, \h{\pi}(g)\, ,  \ee
and the affine commutation relations
\be \label{acr} [\h{q}(f_1),\, \h{q}(f_2)] = 0\, ,\quad [\h{q}(f),\,
\h{\pi}(g)] = i\hbar \h{q}(\bar{f})\, ,\quad [\h{\pi}(g_1),\,
\h{\pi}(g_2)] = i\hbar \h{\pi}(\bar{g})\, , \ee
where
\be \bar{f}^{ab} = f^{c (a}\, g_c{}^{b)}, \quad{\rm and}\quad
     \bar{g}^a{}_b = -\textstyle{\f{1}{2}}\,(g_1^a{}_c\, g_2^c{}_b \,-\,
     g_2^a{}_c\, g_1^c{}_b)\, . \ee

Let us now suppose that there is a diffeomorphism invariant state
$F$ on this $\star$-algebra $\a$. Then, we can repeat the argument
of section \ref{s3.1} to conclude that the cyclic state $\Psi_F$ in
the resulting GNS representation must satisfy:
\be \h{q}(f)|\Psi_F\ket = 0, \quad {\rm and} \quad
\h{\pi}(g)|\Psi_F\ket = 0 \ee
for all test fields $f, g$. But now we cannot make an appeal to the
commutation relations to show a contradiction since the right sides
of (\ref{acr}) also annihilate $\Psi_F$. However since the
representation is cyclic, every vector in $\H$ is the result of
operating on $|\Psi_F\ket$ by some element of $\a$, i.e., a sum of
products of $\h{q}(f)$, $\h\pi(g)$ and $\mathbb{I}$. Hence the
Hilbert space is just 1-dimensional, spanned by $\Psi_F$ and every
element of the $\a$ which is not a multiple of identity annihilates
it. Thus, the representation is trivial. Furthermore, in this
representation $\L(\h{q}_{ab})$ is the zero distribution, whence the
desired positivity requirement is not met.

Again this result does not imply that one cannot construct
interesting representations of the ACR. But there is again a tension
with diffeomorphism invariance, and this may be related to the fact
that the diffeomorphism constraint is not imposed sharply in this
program.

\subsection{Weyl algebra}
\label{s3.3}

Results of the last two sections raise an obvious question: Why is
there a diffeomorphism invariant state that leads to rich quantum
geometry in LQG when there are these no-go results in
geometrodynamics? Where does the key difference lie? The first
essential difference is in the structure of the algebra. While the
gravitational connection $A_a^i$ is a basic canonical variable, the
LQC algebra is generated by their \emph{holonomies}. Originally,
this was motivated by gauge covariance considerations. But then in
the final theory it turned out that there is no operator valued
distribution corresponding to the connection. Indeed, if one were to
construct an Heisenberg algebra $\a$ using operator valued
distributions $\h{A}_a^i, \h{\E}_{ab}^i$, the situation would be
completely analogous to that in the last two sub-sections: one would
again obtain a no go result. Since holonomies are (path ordered)
exponentials of connections, to mimic the situation in LQG, one is
naturally led to the Weyl algebra in geometrodynamics. The second
difference is in the continuity requirement. The existence and
uniqueness result \cite{lost,cf} in LQG suggests that the continuity
assumption on states we imposed in the last two sub-sections is too
strong. I will now show that, once this requirement is appropriately
weakened, the Weyl algebra $\W$ of the \WDW theory does admit a
natural diffeomorphism invariant state which leads to an infinite
dimensional GNS Hilbert space. However, in this representation, it
is not possible to define the momentum operators $\h{p}(h)$; only
their exponentiated versions are well-defined. Therefore, we cannot
use this representation to obtain one for the Heisenberg algebra
$\a$ of the \WDW theory.

Let us first specify the Weyl algebra $\W$. With each pair $(f, h)$
of test fields of section \ref{s3.1}, there is a Weyl operator
$W(f,h)$. Formally, one can think of it as the exponential $W(f,h):=
\exp i (\h{q}(f) + \h{p}(h))$ of the Heisenberg operators and use
(\ref{star1}) and (\ref{ccr}) to ``derive'' the $\star$-operation
and the commutation relations between $W(f,h)$. The result is the
Weyl algebra $\W$. More precisely, elements of $\W$ are finite
linear combinations $\sum_i \lambda_i\, W(f_i, h_i)$ of Weyl
operators $W(f_i, h_i)$, subject to the $\star$-relations
\be \label{star3} W^\star (f, h) = W(-f,-h)\ee
and the analog of the canonical commutation relation
\be \label{product} W(f_1, h_1) W(f_2, h_2) = e^{\f{i\hbar}{2}\int_M
(f_2h_1 - f_1 h_2)\, \dd^3 x}\,\, W(f_1+f_2, h_1+h_2) \ee
which also serves to define products on this vector space.

Does $\W$ admit any diffeomorphism invariant states that lead to
interesting, infinite dimensional GNS representations? As I
mentioned above, we would now like to drop the continuity
requirement guaranteeing the existence of the operator valued
distribution $\h{p}_{ab}(x)$. But since we do wish to explore the
quantum nature of geometry, it is natural to require that the
resulting representation $\L$ should be such that the smeared metric
operators $\L\big(\h{q}(f)\big)$ are well-defined on the GNS Hilbert
space $\H$. This in turn requires that the PLF $F$ is continuous in
$f$ only in the weaker sense that $\lim_{k\rightarrow 0}\,
F[W(kf,h)] = F[W(0,h)]$, where $k \in \R$.

A natural choice is to set
\be \label{plf} F[W(f,h)] := \begin{cases}1 & \text{if}\,\, h=0,\cr
0 & \text{otherwise.}\end{cases} \ee
Diffeomorphism invariance and continuity in $f$ follow by
inspection. That $F$ is a PLF, i.e., that it satisfies $F[A^\star A]
\ge 0$, follows from (\ref{star3}) and (\ref{product}). However, it
is not known if there are other PLFs with these properties and, if
so, whether one can impose additional, well-motivated technical
conditions to narrow down the choice. To my knowledge, a systematic
analysis along the lines of \cite{lost,cf} is yet to be undertaken
in geometrodynamics.

The resulting GNS representation can be described as follows. Recall
that one begins by constructing a vector space $V:= \W/K$, the
quotient of the $\star$-algebra $\W$ by its sub-space $K$ consisting
of elements $A$ satisfying $F(A^\star A) =0$. Thus an element
$\v{A}$ of $V$ is an equivalence class of operators $A\in \W$ where
two are equivalent if their difference lies in $K$. In the present
case, it is easy to check that the equivalence relation on the basic
Weyl operators is given by
\be e^{\f{i\hbar}{2}\int_M f_1h\, \dd^3 x}\,\,W(f_1, h) \approx
e^{\f{i\hbar}{2}\,{\int_M}f_2h\, \dd^3x} \,\, W(f_2, h), \quad
\forall f_1, f_2, h\, . \ee
As a result, the equivalence class to which
$$\underbar{W}(f,h) := e^{\f{i\hbar}{2}\int_M f h\, \dd^3 x}
\,\,W(f, h) $$
belongs is completely characterized by the test field $h_{ab}$.
Therefore, we will denote the ket
$|\overrightarrow{\underbar{W}(f,h)}\ket$ simply by $|h\ket$. Next,
since the inner product is given by $\bra \v{A},\, \v{B} \ket =
F[A^\star B]$, we have
\be \bra h, \, h'\ket = \begin{cases}1 & \text{if}\,\, h=h',\cr 0
& \text{otherwise.}\end{cases} \ee
Thus, $|h\ket$ is an orthonormal basis. The Hilbert space $\H$ is
the Cauchy completion of $V$ with this inner product. Since it
admits an uncountable orthonormal basis, $\H$ is non-separable. But
one can still use the standard Hilbert space machinery, including
the spectral theorem \cite{rudin}. Finally, the representation map
$\L$ is given by:
\be \L\big(W(f,h)\big)\, |h'\ket = e^{\f{i\hbar}{2}\int f^{ab}
h_{ab} \dd^3 x}\,\, e^{i\hbar\int f^{ab} h'_{ab} \dd x}\,\,
|h+h'\ket\, . \ee
The cyclic state is simply
\be |\v{\I}\ket = |\overrightarrow{\underbar{W}(0,0)}\ket = |0\ket
\ee
where, in the last step, we have set $|h=0\ket = |0\ket$. It is easy
to directly verify that the PLF $F$ of (\ref{plf}) is the
expectation value functional in this cyclic state.

Let us explore the salient features of this representation. First,
the Hilbert space is in fact infinite dimensional, so the
representation is non-trivial. Second, since $F[W(f,h)]$ is weakly
continuous in $f$, it follows that the smeared metric operators are
well-defined on $\H$:
\be \L\big(\h{q}(f)\big)\, |h\ket = \big(\textstyle{\int_M
f^{ab}h_{ab}\, \dd^3 x}\big)\,\, |h\ket \, .\ee
Since the state $|h\ket$ is in an eigenstate of the metric operator
$\h{q}_{ab}$ with eigenvalue $h_{ab}(x)$, there is a simple and
natural notion of quantum geometry. However, since the test fields
$h_{ab}$ do not satisfy positivity, \emph{the quantum metric can
have any signature.} In particular, as in section \ref{s3.1}, the
cyclic vector satisfies
\be  \h{q}_{ab}(x)| 0\ket = 0 \, . \ee
Thus, as in LQG, the cyclic state represents the `zero metric'
---just as one would expect of a diffeomorphism invariant state.
This is reminiscent of the situation in 2+1 dimensional gravity
\cite{ew,ahrss} which is exactly soluble.

The key difference from section \ref{s3.1} is that analogous results
do \emph{not} hold for $\h{p}_{ab}$. Indeed, since $F$ is
discontinuous in $h$, the operator $\h{p}_{ab}(h)$ doesn't even
exist; only its exponential, $W(0,h)$ exists. Furthermore, the
exponential has a non-trivial action on the cyclic state:
$\L\big(W(0, h)\big)\, |0\ket = |h\ket$. This is why, in contrast to
the last two sub-sections, the representation is non-trivial. But
since it does not admit operators $\h{p}(h)$, it cannot be used to
construct a representation of the canonical $\star$-algebra $\a$.

It is instructive to compare this version of the \WDW theory with
LQG. In both cases, the GNS representation is non-trivial. The no go
results of sections \ref{s3.1} and \ref{s3.2} are bypassed because
there are no operators corresponding to the variable canonically
conjugate to geometry (the connection $A_a^i(x)$ in LQG and the
momentum $p^{ab}$ in geometrodynamics.) Put differently, in both
cases, interesting background independent kinematics can exist only
if we drop the requirement that these operators should exist:
Intuition derived from our experience with Minkowskian quantum field
theories fails to carry over once in the diffeomorphism covariant
context. But the exponentials of these variables do lead to
well-defined operators on the GNS Hilbert space. They act
non-trivially on the cyclic state, producing eigenstates of quantum
geometry. In both cases, the GNS Hilbert space $\H$ is
non-separable. In the \WDW case, we saw that this feature arises
because the elementary excitations of geometry $|h\ket$ cerated by
operating on the cyclic state $|0\ket$ by $W(0,h) = e^{i\h{p}(h)}$
are \emph{orthogonal} for distinct test fields $h$. This is also a
direct consequence of background independence: there is no non-zero
diffeomorphism covariant expression we can write for $\bra h,\,
h'\ket$ if $h \not= h'$ and obtain an infinite dimensional Hilbert
space. In LQG the holonomy operator replaces $e^{i\h{p}(h)}$ but
after that the situation is completely analogous. In this sense the
non-separability of the kinematical Hilbert space can also be traced
back to background independence.%
\footnote{In LQG the \emph{physical} Hilbert space will be separable
because the imposition of constraints factors out the redundancy in
the kinematical framework \cite{alrev,crbook}. This idea is realized
in detail in LQC \cite{aa-badhonef}.}

In spite of these similarities, the nature of quantum geometry in
the two theories is very different. In LQG, the eigenstates of
geometrical operators are spin-network states which have support on
\emph{1-dimensional} graphs. These elementary excitations are again
generated by the action of exponentiated operators on the cyclic
state but now the exponentiated operators are holonomies which have
support on 1-dimensional curves (which constitute the edges of
graphs). Thus, quantum geometry is \emph{polymer-like}. As a direct
consequence, geometrical operators on $\H$ like areas of surfaces
and volumes of regions have purely discrete eigenvalues. This
discreteness has important consequences in the black hole entropy
calculations \cite{entropy} as well as in the dynamics of quantum
cosmology \cite{aa-badhonef}. By contrast, quantum geometries
constructed in this section using the \WDW theory are much more
``tame.'' They are created by the action of $\exp i\h{p}(h)$ where
the smearing fields have 3-dimensional support. If one were to
construct geometric operators in this theory, they would have
continuous eigenvalues but with any sign. Therefore it is not
obvious why quantum horizons would have only a finite number of
surface states. If one were interested in addressing such issues in
the present version of the \WDW theory, one would need brand new
ideas.

\section{Discussion}
\label{s4}

Familiar systems with a finite number of degrees of freedom often
admit a linear phase space, coordinatized by configuration and
momentum variables $q^i, p_i$. One can then readily introduce Weyl
operators $W(\lambda_i, \mu^i)$ which generate a $\star$-algebra
$\W$. The task of finding a representation of this algebra is
rendered trivial by von Neumann's uniqueness theorem: up to
isomorphism, there is a unique irreducible representation of $\W$ by
unitary operators $\L(W(\lambda_i,\mu^i))$ on a Hilbert space in
which the $\L(W(\lambda_i,\mu^i))$ are weakly continuous in
$\lambda_i, \mu^i$. The weak continuity requirement is necessary and
sufficient for the existence of the configuration and momentum
operators $q^i,\,p_i$, whence it is natural to impose it in standard
quantum mechanics. Thanks to the von-Neumann theorem, we routinely
restrict ourselves to the Schr\"odinger representation in
non-relativistic quantum mechanics.

For systems with an infinite number of degrees of freedom, the
situation is very different because the Weyl algebras admit an
infinite number of unitarily inequivalent representations
\cite{gw,rh,ge}. It is therefore a pleasant surprise that
Lewandowski, Okolow, Sahlmann and Thiemann \cite{lost} and
Fleischhack \cite{cf} could single out a canonical representation of
the holonomy-flux algebra of LQG by making an appeal to background
independence. Their result is somewhat analogous to the uniqueness
of the Poincar\'e invariant vacuum in Minkowskian quantum field
theories \cite{unique}. However, the Minkowskian uniqueness theorem
assumes free field dynamics. By contrast, the result of
\cite{lost,cf} is purely kinematical; it is applicable to \emph{any}
diffeomorphism covariant
theory with a holonomy-flux $\star$-algebra.%
\footnote{The uniqueness theorems of \cite{lost,cf} also involve a
number of highly non-trivial mathematical subtleties. In particular,
restriction to the piecewise analytical category plays an important
role in constructions and proofs. More generally, the detailed
arguments require a much higher degree of mathematical
sophistication than in Minkowskian free field theories.}
Thus the requirement of diffeomorphism invariance on a state is much
more powerful than that of Poincar\'e invariance. The GNS
representation obtained from this unique state provides the quantum
kinematics in LQG. This kinematics has two non-standard features:
The Hilbert space is non-separable (see, however, footnote 5) and it
does not admit a local connection operator(valued distribution)
$A_a^i(x)$. 
As we saw in section \ref{s3.3}, both these features can
be traced back to the diffeomorphism invariance.

But returning to the basics, one might ask: Why look for a state
which is diffeomorphism invariant in the first place? The reason is
that it naturally leads to a kinematic setting which is
well-tailored for the task of imposing the diffeomorphism
constraint. Recall that, in Dirac or BRST quantization of general
constrained systems, one first constructs a kinematical Hilbert
space on which the group generated by constraints is unitarily
implemented. One then uses a standard ``group averaging procedure''
to construct physical states (which are left invariant by this
group) and to introduce an inner product thereon \cite{dm,almmt}.
Now, as we saw in section \ref{s2}, given any diffeomorphism
invariant state $F$ on the LQG $\star$-algebra the diffeomorphism
group is automatically represented by unitary transformations on the
resulting GNS Hilbert space $\H$. The kinematical setup is thus
already ``prepared'' for the group averaging procedure. Indeed, this
is how the diffeomorphism constraint of general relativity is
implemented in LQG \cite{almmt,alrev,crbook,ttbook}. Thus, the
availability of a diffeomorphism invariant state $F$ in LQG provides
a ``royal road'' to the imposition of the diffeomorphism constraint.
The Hilbert space $\H_{\rm diff}$ of diffeomorphism invariant states
is again infinite dimensional: There are indeed infinitely many
diffeomorphism invariant states but on the algebra of
\emph{diffeomorphism invariant operators} (such as the volume of
$M$). These states do not descend down to the kinematical algebra
$\a$ or $W$ (because the holonomy and triad-flux operators are not
diffeomorphism invariant).

As pointed out in section \ref{s1}, at first the uniqueness of a
diffeomorphism invariant state on the LQG $\star$-algebras seems
surprising from the perspective of geometrodynamics. Therefore in
section \ref{s3} we analyzed diffeomorphism invariant states on the
geometrodynamical $\star$-algebras. We found that the canonical
$\star$-algebra $\a$ used there does not admit even a single
diffeomorphism invariant state. This is in sharp contrast to the
general expectation that there should be infinitely many states
$\Psi(q)$ built, e.g., from integrals of curvature invariants. It
has often been implicitly assumed that such $\Psi(q)$ would be
square integrable with respect to some diffeomorphism invariant
measure on the space of 3-metrics (or an appropriate extension
thereof) and elements of $\a$ would have well-defined action on
$\Psi(q)$. If this were true for even a single such $\Psi(q)$, via
expectation values it would have defined a diffeomorphism invariant
state $F$ on the canonical $\star$-algebra. Non existence of such an
$F$ shows that the heuristic expectation is flawed. For the affine
algebra, there is an unique 
diffeomorphism invariant state but the resulting GNS Hilbert space
is just 1-dimensional. So, the GNS representation would be a poor
candidate for quantum kinematics. The situation is very different
for the Weyl algebra $\W$ of geometrodynamics. This algebra does
admit a non-trivial diffeomorphism invariant state. In the resulting
GNS representation, the Hilbert space is infinite dimensional and,
furthermore, quantum geometry is well-defined. Symmetry reduced
versions of this representation have been used in the context of
cosmology and gravitational collapse (see e.g. \cite{vh}). The GNS
representation has the same non-standard features we encountered in
LQG: the Hilbert space $\H$ is non-separable and does not carry a
representation of the momentum operators $\h{p}(h)$; only their
exponentials are well-defined. As a consequence, it does not lead to
a representation of the canonical $\star$-algebra $\a$ (just as the
unique diffeomorphism invariant state on the LQG $\star$-algebra
does not lead to a representation of the canonical algebra generated
by $A_a^i,\, \E_{ab}^i$). In spite of these similarities, however,
quantum geometry in this representation is very different from that
in LQG. The basic excitations are 3-dimensional rather than
1-dimensional ``polymers'', and it is unlikely that the geometrical
operators will have discrete eigenvalues.

Let us conclude with a comparison between gauge and diffeomorphism
invariance. Consider the free Maxwell theory in Minkowski space. The
canonical $\star$-algebra $\a$ is generated by operator valued
distributions $A_a(x), \, E^a(x)$. We can decompose them into
transverse and longitudinal parts and the two decouple. The gauge
transformation $A_a \rightarrow A_a + \dd\alpha$ gives rise to an
automorphism $\theta_\alpha$ on $\a$ which has a non-trivial action
only on the longitudinal operators $A^{\rm L}_a(x)$. The reasoning
of section \ref{s3} implies that, contrary to one's initial
expectations, there is \emph{no} state $F$ on $\a$ which is
invariant under these gauge automorphisms. What about the Weyl
algebra $\W$ generated by the exponentials  $W(f,g)= e^{i(\h{A}(f) +
\h{E}(g))}$? Now, there are \emph{infinitely many} gauge invariant
states. On the transverse modes we can use any state from the
standard Fock representation and on the longitudinal modes we can
use a ``polymer state'' as in section \ref{s3.3}. (For explicit
expressions, see \cite{aa-ilqg}.) The resulting GNS Hilbert space
$\H$ will admit concrete operators $\h{A}(f_T), \, \h{E}(g^T)$ and
$\h{E}(g^L)$, where the descriptors $T,\,L$ denote transverse and
longitudinal test fields. However, $\H$ will not carry connection
operators $\h{A}(f_L)$ smeared with longitudinal test fields; only
the exponentials of these operators would be well-defined.%
\footnote{This general idea was in fact used by Thirring and
Narnhofer \cite{thirring} to show that there is a covariant
quantization procedure that avoids the introduction of an indefinite
metric used in the Gupta-Bleuler quantization procedure one finds in
text books.}
%
Because the gauge transformations are unitarily represented on $\H$,
we can carry out group averaging to obtain the physical Hilbert
space $\H_{\rm phys}$. Physical states are functionals only of
transverse modes and the $\H_{\rm phys}$ is separable. Thus there
are similarities with diffeomorphism invariance. However, there is
also a key difference: While the Weyl algebra of Maxwell fields
admits infinitely many gauge invariant states, the holonomy-flux
algebra of LQG admits a \emph{unique} diffeomorphism invariant
state. In this sense, requirement of diffeomorphism invariance is
vastly more powerful.

\section*{Acknowledgments}
The need for clarification of the unusually strong role of
diffeomorphism invariance was brought to forefront during
discussions in the 2007 KITP workshop on Classical Singularities in
Quantum Space-times. I would like to thank Ted Jacobson and Don
Marolf for raising interesting questions and Jerzy Lewandowski for
discussions. This work was supported in part by the NSF grants
PHY0456913 and PHY0854743, The George A. and Margaret M. Downsbrough
Endowment and the Eberly research funds of Penn State.


\begin{thebibliography}{99}


\bibitem{lost} J.~Lewandowski, A.~Okolow, H.~Sahlmann and
    T.~Thiemann, {Uniqueness of diffeomorphism invariant states on
    holonomy flux algebras}, Comm. Math. Phys. \textbf{267}, 703-733
   (2006).

\bibitem{cf} C.~Fleishchack, {Representations of the Weyl algebra in
    quantum geometry}, Commun. Math. Phys. \textbf{285}, 67-140
    (2009).

\bibitem{gw}L.~Garding and A.~S.~Wightman, Proc. Nat.
    Acad. Sci. U.S.A., \textbf{40}, 622-626 (1954).

\bibitem{rh}R.~Haag, On quantum field theories, Danske Vid. Selsk.
    Mat.-fys. Medd. \textbf{29}, No.12 (1955).

\bibitem{ge}G.~M.~Emch, \emph{Algebraic methods in statistical
    mechanics and quantum field theory}, (Wiley-Interscience, New
    York, 1972).

\bibitem{alrev} A.~Ashtekar and J.~Lewandowski, {Background
    independent quantum gravity: A status report}, Class. Quant.
    Grav. {\bf 21}, R53-R152 (2004).

\bibitem{crbook} C.~Rovelli,{\em Quantum Gravity}. (Cambridge
    University Press, Cambridge (2004)).

\bibitem{ttbook} T.~Thiemann, {\em Introduction to Modern
    Canonical Quantum General Relativity.} (Cambridge University Press,
    Cambridge, (2007)).

\bibitem{almmt} A.~Ashtekar, J.~Lewandowski, D.~Marolf, J.~Mour\~ao
    and T.~Thiemann, Quantization of diffeomorphism invariant
    theories of connections with local degrees of freedom. {Jour.
    Math. Phys.} \textbf{36}, 6456-6493 (1995).

\bibitem{rs} R.~Rovelli and L.~Smolin, Discreteness of area and
    volume in quantum gravity, Nucl. Phys. \textbf{B442}593-622
    (1995); Erratum \textbf{B456}, 753 (1996).

\bibitem{al5} A.~Ashtekar and J.~Lewandowski, Quantum theory of
    geometry I: Area operators, Class. Quantum Grav. \textbf{14},
    A55-81 (1997).

\bibitem{alvol} A.~Ashtekar and J.~Lewandowski, Quantum theory of
geometry II: Volume operators, Adv. Theo. \& Math. Phys.
    \textbf{1}, 388-429 (1997).

\bibitem{acz} A.~Ashtekar, A.~Corichi and J.~A.~Zapata, Quantum
    Theory of Geometry III: Non-commutativity of
    Riemannian Structures, Class. Quant. Grav. \textbf{15}, 2955-2972
    (1998).

\bibitem{tt} T.~Thiemann, A length operator for canonical quantum gravity
J. Math. Phys. \textbf{39}, 3372-3392 (1998).

\bibitem{jw} J.~A.~Wheeler, Superspace and quantum
    geometrodynamics, in: Battelle Rencontres, edited by
    J.~A.~Wheeler and C.~M.~DeWitt (W.~A.~Benjamin, New York, 1972).

\bibitem{kk} A.~Komar, Quantization program for general relativity.
    In: \textit{Relativity}, Carmeli, M., Fickler, S.~I.,
Witten, L.\ (eds.) (Plenum, New York (1970);\\
K.~Kucha$\check{\rm r}$, Canonical methods of quantization. In:
\textit{Quantum Gravity 2, A Second Oxford Symposium}, Isham, C.~J.,
Penrose, R., Sciama, D.~W.\ (eds.) (Clarendon Press, Oxford (1981)).

\bibitem{jk}J.~Klauder,Soluble models of quantum gravitation, in
\emph{Relativity}, Carmeli, M., Fickler, S.~I., Witten, L.\ (eds.)
(Plenum, New York (1970);\\
Fundamentals of quantum gravity, J. Phys. Conf. Ser. \textbf{87}
012012 (2007).

\bibitem{dm} D.~Marolf, Refined algebraic quantization:
Systems with a single constraint. \texttt{arXive:gr-qc/9508015};\\
{Quantum observables and recollapsing dynamics}. Class. Quant. Grav.
{\bf 12}, 1199--1220 (1994);\\
A.~Ashtekar, L.~Bombelli and A.~Corichi, Semiclassical states for
constrained systems, Phys. Rev. \textbf{D72}, 025008 (2005)

\bibitem{ew}E.~Witten, 2+1 dimensional gravity as an exactly
soluble system, Nucl. Phys. \textbf{B311}, 46-78 (1988).

\bibitem{ahrss} A.~Ashtekar, V.~Husain, C.~Rovelli, J.~Samuel and
L.~Smolin, 2+1 gravity as a toy model for the 3+1 theory, Class.
Quant. Hrav. \textbf{6}, L185-L193 (1989).

\bibitem{rudin}W.~Rudin, \emph{Functional Analysis}, (McGraw Hill,
New York, 1973).

\bibitem{aa-badhonef} A.~Ashtekar, Loop Quantum Cosmology: An
    Overview (2007), Gen. Rel. and Grav. \textbf{41}, 707-741
    (2009).

\bibitem{entropy} A.~Ashtekar, J.~Baez, A.~Corichi and
    K.~Krasnov, Quantum Geometry and Black Hole Entropy, Phys. Rev. Lett.
    \textbf{80}, 904-907 (1998).\\
    A.~Ashtekar, J.~Baez, and K.~Krasnov, Quantum Geometry
    of Isolated Horizons and Black Hole Entropy, Adv. Theor. Math.
    Phys. \textbf{4}, 1-94 (2000).\\
    A.~Ashtekar, J.~Engle and C.~Van Den Broeck, Class. Quant. Grav.

\bibitem{unique} I. E. Segal, Mathematical characterization of the
    physical vacuum, Illinois J. Math. \textbf{6}, 500-523 (1962).

\bibitem{vh}V.~Husain and O.~Winkler, Semiclassical states for
    quantum cosmology, Phys. Rev. \textbf{D75} 024014 (2007).\\
    V.~Husain, Critical behavior in gravitational collapse,
    \texttt{arXiv:0808.0949}.

\bibitem{thirring}W.~Thirring and H.~Narnhofer, Covariant QED
    without indefinite metric, Rev. Math. Phys. \textbf{SI1},
    197-211 (1992).

\bibitem{aa-ilqg}A.~Ashtekar, International loop quantum gravity
    seminar, http://relativity.phys.lsu.edu/ilqgs/ashtekar022707.pdf



\end{thebibliography}
\end{document}